# Applications of Social Media in Hydroinformatics: A Survey

**Yufeng Yu, Yuelong Zhu, Dingsheng Wan,Qun Zhao**     hhuheiyun@126.COM

*College of Computer and Information*
*Hohai University*
*Nanjing, Jiangsu, China*

**Kai Shu, Huan Liu**     huan.liu@asu.edu

*School of Computing, Informatics, and Decision Systems Engineering*
*Arizona State University*
*Tempe, Arizona, U.S.A*

**Abstract**

Floods of research and practical applications employ social media data for a wide range of public applications, including environmental monitoring, water resource managing, disaster and emergency response, etc. Hydroinformatics can benefit from the social media technologies with newly emerged data, techniques and analytical tools to handle large datasets, from which creative ideas and new values could be mined. This paper first proposes a 4W (What, Why, When, hoW) model and a methodological structure to better understand and represent the application of social media to hydroinformatics, then provides an overview of academic research of applying social media to hydroinformatics such as water environment, water resources, flood, drought and water Scarcity management. At last，some advanced topics and suggestions of water-related social media applications from data collection, data quality management, fake news detection, privacy issues , algorithms and platforms  was present to hydroinformatics managers and researchers based on previous discussion.

**Keywords:** Social Media, Big Data, Hydroinformatics, Social Media Mining, Water Resource, Data Quality, Fake News

## 1 Introduction

In the past two decades, breakthrough in remote sensing technology has brought unparalleled advances in earth environmental observation (EO). New satellite, space, airborne, shipborne and ground-based remote sensing systems are springing up all over the world [1,2], which lead to the huge number of EO datasets such as geographic data, meteorological data and environmental monitoring data was acquired and generated from low to high spatial, temporal and radiometric resolution at a breathless pace, both in size and variety [3].It goes without saying that a big data era has been boosted in the field of EO, which will bring great opportunity as well as great challenges to both scientists and information technology experts[4].

As a critical part of the environmental observation, water resource is the source of life in all things in the world. It is the fundamental requirement for health and the main need for

YUFENG YU, YUELONG ZHU, DINGSHENG WAN, QUN ZHAO, KAI SHU AND HUAN LIU

industrialization. It is mankind's most precious commodity and its sustainable development and utilization are directly related to the national economy and people's livelihood. However, the currently ongoing rapid development of economy and technology tremendously alters our environment and has significant altered hydrological processes, causing a great variety of water issues (water pollution, flood, drought and water scarcity), forming a serious threat to society development, slowing economic growth, threatening community health. Therefore, how to apply the latest information technology to control and manage water-related issues, minimize damage for life and property, and maximize usage benefits efficiency becomes a critical issue for water and environmental research domains.

The majority of existing approaches utilize the systems and algorithms to record water-related information from a remote location, remote sensing and Internet of Things (IoT) play an essential role in many applications of water resource, such as water environment monitoring; water resource monitoring, managing and controlling; water crisis and emergency response, etc.[5-6]. However, remote sensing and IoT data may not be always available, especially in situations of crisis management. The apparent overabundance of data is often accompanied by a simultaneous "information dearth": a lack of information may arise because sensors are not available for certain regions or the number of available sensors is not enough to cover the territory with a suitable resolution. In hydrology, this problem is attributed to the so-called "ungauged" or "poorly gauged" catchments [8]. In response, other sources of data, such as social media[7] are emerging that provides important information and can supplement traditional sensors. These sources include data provided by people directly linked to affected areas or associated areas, which can be used in many real-time water-related scenarios such as flood risk, water crisis and assist in water resources management [9-10].

Built on the ideological and technological foundations of Web 2.0, social media can guide and offer incentives to users to share information and communicate through real-time online communication via computers or mobile devices. It plays a relevant role in our daily lives and provides a unique opportunity to gain valuable insight on information flow and social networking in our society. Social media has several major functions in water-related management processes, including one- and two-way information sharing, situational awareness, rumor control, reconnection, and decision making. Clear evidence suggests that social media is increasingly used as a dissemination and communication tool for water-related business such as flood forecasting and assessment, water management, and water crisis management. However, further development, validation, and implementation of viable and accurate water-related management requires a more detailed mapping and understanding of the evolving water as well as the efficiency and capabilities of hydrodynamic modeling frameworks. Therefore, how to find the best way to extract meaningful information from social media and integrate this information with data from other sources to achieve greater reliability, how to ensure this information to be useful for hydrological models to support decision-making with regard to water-related design and management, are still multiple challenges for applying social media to hydroinformatics.

This paper provides an overview of academic research related to a link between social media and hydroinformatics. In addition, it contributes to the understanding and construction of the

# APPLICATIONS OF SOCIAL MEDIA IN HYDROINFORMATICS: A SURVEY

data-driven hydrological modeling such as flood prediction, water resources monitoring, water environment monitoring and water crisis response between social media and hydroinformatics. The remainder of our paper proceeds as follows: first we status quo of the literature on social media and the highlight the status of our article afterwards. Second, we propose a 4W (What, Why, When, hoW) model and a methodological structure to better understand and represent the application of social media to hydroinformatics. Third systematically summarize the applications and research methods of social media in flood forecasting and management, water resources monitoring and management, water environment monitoring and inclusion and water crisis response. And then, we present some future research directions for the development of water-related social media from different aspects to hydroinformatics managers and researchers. Finally, we conclude our article.

## 2 Literature Background

This section gives the basic concepts of social media and its mining tasks, then briefly describes the main application areas of social media, and finally discusses the current status of Hydroinformat -ics.

### 2.1 Social Media

Social media [11] is a group of web-based and mobile-based Internet applications established on the conceptual and technical foundations of Web 2.0 and allow the publishing, sharing and distributing of user-generated content. It is conglomerate of different types of social media sites including traditional media such as newspaper, radio, television and nontraditional media such as Twitter, Facebook, Weibo, and Weichat (the popular Chinese version of Twitter) etc. Social media can provide the users a convenient-to-use way to learn, communicate and share information with each other on an unprecedented scale and unseen rates than traditional media. The popularity of social media continues to grow exponentially, leading to a fundamental evolution of social networks, blogs, social bookmarking applications, social news, media (text, photo, audio, and video) sharing, product and business review sites, etc. Facebook[1] and Wechat[2], most popular social networking site in US and China, recorded more than 2.2 billion and 1.06 billion active users as of 2nd quarter 2018 respectively. This number suggests that the users owned by Facebook and WeChat are basically the same as the population of the United States has and more than the population of any continent except Asia.

    Floods of user-generated content are created, shared and disseminated on social media sites every day. This trend is more likely to be continued in a faster, deeper form in the future. Hence, it is critical for our users to make sure how to obtain, manage and utility information their need from massive user-generated data. According to [12], Social media growth is driven by three challenges: (1) how does a user's opinion be correct expressed and be heard? (2) Which source of information should a user identify and use? (3) How can user experience be improved? This presents ample opportunities and challenges for researches to develop new data mining algorithms and methods to

---

[1] https://www.statista.com/statistics/264810/number-of-monthly-active-facebook-users-worldwide/
[2] https://www.statista.com/statistics/255778/number-of-active-wechat-messenger-accounts/

mine information hidden in the social media data to answer aforementioned questions.

**2.2 Social Media Mining**

Social media data are essentially different from conventional attribute-value data for traditional data mining.

Apart from enormous size, social media data are largely user-generated content on social media sites and have noisy, distributed, unstructured and dynamic characters with abundant social relations such as friendships and followers-followees. This new type of data mandates new computational data analysis approaches that can combine social theories with statistical and data mining methods. The fast-growing interests and intensifying need to harness social media data demand for new techniques ushers in and entails a new interdisciplinary field–social media mining.

Aiming to combine, extend, and adapt methods for the analysis of social media data [13], social media mining is a rapidly growing new interdisciplinary field at the crossroad of disparate disciplines deeply rooted in computer science and social sciences. Academically speaking, social media mining is the process of representing, analyzing, and extracting actionable patterns from social media data. Social media mining grows a new kind of data science research field in which we can develop social and computational theories, to analyze recalcitrant social media data, and to help bridge the gap from what we know (social and computational theories) to what we want to know about the vast social media world with computational tools.

Social media mining is an emerging field where there are more problems than ready solutions. Mining social media data is the task of mining user-generated content with social relations. There are some representative research issues in mining social media [14].

1) Sentiment analysis. Sentiment analysis(or opinion mining) [15-16], is the field of study that analyzes people's opinions, sentiments, evaluations, appraisals, attitudes, and emotions towards entities such as organizations, services, products, individuals, issues, events, topics, and their attributes. With the explosive growth of social media (i.e., reviews, forum discussions, blogs and social networks) on the Web, individuals and organizations are increasingly using public opinions in these media for their decision making. However, finding and monitoring opinion sites on the Web and distilling the information contained in them remains a formidable task on the reason that: 1) monitoring opinions related to a particular environmental issue on social media sites is a new challenge, 2) languages used to create contents are ambiguous, and 3) lack of ground truth to performance evaluation of sentiment analysis.

2) Social similarity analysis. Social forces connect individuals in different ways. Social similarity is a term used to measure degree of assortativity in their connectivity networks when individuals get connected. Influence [17-18] and homophily [19] are common forces to represent social similarity and both of them give rise to assortative networks recently. Hence, how to measure and model influence and homophily to reveal the laws of individuals interaction in social media, is one of the main tasks of social media mining. Moreover, it is important to know whether the underlying social network is influence driven or homophily driven because influence makes "friends become similar" while homophily makes "similar individuals become friends". However, distinguishing homophily and influence is a challenge task because of that most social network has a mixture of both [14, 20].

3) Social Recommendation. Individuals in social media make a variety of decisions such as purchasing a service, buying a product, adding a friend, and renting a movie, among others, on a daily basis. Recommender systems are applications and algorithms developed to recommend products that would be interesting to individuals. Traditional recommendation systems [21-22] attempt to recommend items based on aggregated ratings of objects from users or past purchase histories of users. Social recommendation systems are based on the hypothesis that people who are socially connected are more likely to share the same or similar interests (homophily).In addition to the traditional recommendation means, they also make use of user's social network and related information. Therefore, users can be easily influenced by their trust friends and keen to their friends' recommendations [23]. Recommendation systems are designed to recommend individual-based choices. Thus, the same query issued by different individuals should result in different recommendations. As simple as this process may look, a social recommendation system actually has to deal with challenges such as cold-start problem, data sparsity, recommendation attacks and individuals' privacy [14].

4) Information Diffusion and Provenance. Society provides means for individuals to exchange information through various channels, but different research fields may have different views on what is the process of information dissemination. Thus, information diffusion can be defined as the process by which a piece of information (knowledge) is spread and reaches individuals through interactions [14]. Meanwhile, information provenance provides a similar value to its users and can be considered as the origins, custody, and ownership of a piece of information published in a social media setting [24]. Researchers study how information diffuses and explore different models of information diffusion to analyze the spread of rumors, computer viruses, and diseases during outbreaks. According to the distributed and dynamic characteristics of social media data, (1) how information spreads in a social media network and which factors affect the spread(information diffusion), and (2) what plausible sources are (information provenance) , are two important research issues from the social media viewpoint and recognized as key issues to differentiate rumors from truth[12].

5) Privacy and Security. The easy access and widespread use of social media raise concerns about user privacy and security issues. In the social media era, people would like to have as many friends and information as possible while to be as private and security as possible when necessary; and social networking sites need to encourage users to easily find each other and expand their friendship network as widely as possible to meet their business development requirements; these will poses new security opportunities and challenges to fend off security threats to users and organizations. With continuous application and popularization of social media ecosystems such as new scenes, new platforms, and new applications, social media has been the target of numerous passive as well as active attacks and individuals may put themselves and members of their social networks at risk for a variety of attacks including Identity theft, Spam attack, Malware attacks, Sybil attacks, Social phishing, Impersonation, Hijacking and the like [25]. Therefore, how to 1) enhance security control mechanism for social network [26], 2) protect the personal information disclosure and privacy for social media users [27] and 3) manage digital rights for multimedia content [28] become practical application problems of social media mining.

**2.3 Social Media Based Applications**

The social media mining techniques and methods can be applied to social media data sources for discovering relevant knowledge that can be used to improve the decision making of individual users and companies in different domains, such as marketing, customer relationship management (CRM), knowledge sharing, user experiences -based visualization, emergency response and so on [29].

Marketing researchers believe that social media and cloud computing as a marketing tool and as an integral part of the integrated marketing communication strategies of firms can offer a unique opportunity for businesses to obtain opinions and user intelligence from a vast number of customers, generating more targeted advertising and marketing campaigns [30]. On this basis, many studies gave different viewer of social media marketing, including consumer attitude and behavior [31], marketing impact of customer communication and recommendation [32] and branding issues in the social media environment [33].

CRM is a mechanism to manage the interactions of a company with its current and potential customers. At present, numerous companies have adopted social media to manage and improve their relationships, such as customer experiences [34], relationship quality and customer satisfaction [35], customer knowledge management (CKM) and trust cultivation [36], with their customers.

Knowledge sharing is an activity in which individuals, friends, families, communities, and organizations exchange information, skills, or expertise [37]. Social media contribute and facilitate knowledge sharing in online communities, particularly knowledge related to product information, travel information, and/or customer experiences. Currently, Social media knowledge sharing focuses on the reasons why users are keen to knowledge sharing [38], the motivations and determinant factors of knowledge sharing[39] and the impacts of knowledge sharing in the social media environment[40].

Big data from social media needs to be visualized for better user experiences and services. For example, the large volume of numerical data (usually in tabular form) can be transformed into different formats. Consequently, user understandability can be increased. The capability of supporting timely decisions based on visualizing such big data is essential to various domains, e.g., business success, clinical treatments, cyber and national security, and disaster management [41]. Thus, user-experience-based visualization has been regarded as important for supporting decision makers in making better decisions. More particularly, visualization is also regarded as a crucial data analytic tool for social media [42] because it can better understand users' needs in social networking services.

As a sudden, urgent, usually unexpected incident or occurrence that requires an immediate reaction or assistance, emergency event becomes a common phenomenon in the daily life of the public, such as flood, fire, storms, traffic congestion, and so on. Once emergency event occurred,

# APPLICATIONS OF SOCIAL MEDIA IN HYDROINFORMATICS: A SURVEY

decision-making or public service departments need to respond or assist quickly to protect the lives and property of the public. Therefore, the use of social media mining technology to quickly detect, resistant, and analyze real-time emergencies has received more and more research attention. Based on the assumption that Twitter is a distributed sensor system and the real-time, spatial and temporal features of Twitter, the messages from twitter users were collected and used to emergency planning, risk and damage assessment activities through earthquake detection[43], forest fire detection[44], flash flood warning[45] and rapid inundation mapping[46].

In addition to the above-mentioned disciplines, social media applications can be found in many other areas. Relevant research results can be found in medical sector [47], public relationships (PR) sector [48] and tourism industry [49].

Unfortunately, most work to date has focused on Twitter or Facebooks, emphasizing related topics of major concern in the United States, with little work concerning issues in other countries. Moreover, automatically detecting and mining real-time, valuable information from social media is not that easy. The potential challenge is summarized as follow.

(1)The data volume of all social media users is up to TB level every day. Moreover, the data volume is still growing at an alarming rate every day. Thus, how to organize, store, and quickly index these social media big data is a huge challenge and difficult task.

(2) Unlike physical sensors, social sensors are activated by specific events, so the data collected by it are noisier and more redundant than that collected by real sensors. That is, when a social network user makes a poster about a special event, he/she can be considered as a social sensor for this event. But these social sensors may post, spread and forward some incorrect, incomplete or fake information for subjective or objective reasons.

(3) Social media data usually has high value and high dimensional characteristics. But the phenomenon of "high volume, low value" from the big data area also exists in the social media data. Therefore, How to extract valuable and meaningful information from the huge volume of social big data is a challenge for social media data mining.

(4) Social media device have fast data input/output capabilities. That is，the velocity of collecting social media data is faster than that of processing and analyzing them，which brings the big challenges for processing and analyzing social media data.

**2.4 Status of Hydroinformatics**

Hydroinformatics, originated from the computational hydraulics [50], is one interdisciplinary field of technology which focuses on integrating information and communication technologies (ICTs) with hydrologic, hydraulic, environmental science and engineering to address the increasingly serious problems of the equitable and efficient use of water for different purposes. The two main lines of hydroinformatics, data mining for knowledge discovery and knowledge management [51], are strongly dependent on information of which data, both textual or non-textual, is the major carrier. Data from smart meters, smart sensors, remote sensing, crowdsourcing, earth observation systems, etc., will prompt hydroinformatics into the inevitable social media big data era.

Hydroinformatics comprises many state-of-the-art applications of modern information technologies in water management and decision making. It focuses on [52]:

• New themes such as computational intelligence, control systems, and their application in data-driven hydrological modeling,

• Optimization and real-time control of models

• Flood modeling for management of module integrating modeling theory, hydraulics, and flood simulation

• Water resources modeling for water resource scheduling and allocation, water environment management and water resources protection

• Decision support systems module integrating system analysis, decision support system theory, and model integration

Over the last 20 years the hydroinformatics has shown its capabilities to address some of water-related issues in a way that it can meaningfully provide integration between data, models and decision support. However, the current practical and research effort is still very much technocratic (or techno-centric) which in turn may restrict the potential of hydroinformatics in its scope and its reach [53]. Furthermore, many researchers working in hydroinformatics are still struggling to get full-scale acceptance within the hydrological community, which is dominated by larger groups of traditionalists who care less about data-driven model and more about physics. With the continuous development of ICTs and the updating of hydrological models, hydroinformatics confronts the following challenges:

• With the extension of hydroinformatics onto the sociotechnical dimension, the primary role of hydroinformatics nowadays is in the development and installation of sociotechnical arrangements that can truly enable the right balance between quantities (i.e., measurable substance, matter, structure) and qualities (i.e., patterns, dependences, interrelationships, contexts, perceptions, feelings, emotions, subjective experiences, etc.) and apply them meaningfully in our research and practice. Hence, the traditional perception of hydroinformatics has to change into one where ideas emerge from qualities and social needs and concerns and proceeds through indefinable feedback cycles where the acceptable social, ethical, technical and environmental norms and standards continuously change, leading to a better understanding of phenomena and better interventions into the physical environment.

• As one important component of the natural environment, water-related information is the interaction bridge between the water environment and human society. Therefore, there is an urgent need to develop the global monitoring information system on water to provide wide spectrum of information from the local level and up to national and global levels (e.g., monitoring data, public documents, comprehensive national plans, available and appropriate technologies) for water management and to monitor progress against targets. But it is more than a matter of better sensors and more satellites. There need to be corresponding improvements in ground-based monitoring networks, and an integration of knowledge from all sources, including complementary airborne monitoring systems in order to improve water resources management [54].

• Data is the foundation of all water related businesses such as modeling, analysis, planning, and management. According Information Handling Services (IHS) Markit, there will be more than 4 million smart meters, smart sensors and smart services, remote sensing, etc., has been set up by

# APPLICATIONS OF SOCIAL MEDIA IN HYDROINFORMATICS: A SURVEY

2022 to collect water-related information, which will prompt hydroinformatics into the inevitable big data era. Therefore, how to use computer techniques and tools to address the problems related to data capture, storage, searching, sharing, analysis, management and visualization is a big challenge. Moreover, the 5V characteristic of big data makes it difficult to discover the relevant topics, trends and events such as climate change, flood and drought management, the global water cycle, the interaction between water environment and human society, water resource management and future water planning in dynamic big data.

• The future of hydroinformatics is directed at supporting and indeed enabling holistic analysis, design, installation and development of on-line and real-time construction, operation and management of water systems that will be highly adaptive to changing conditions, such as those that may occur slowly over years (e.g., climate change effects) and over a few hours (e.g., flood conditions), or in extreme cases even over some minutes (e.g., evacuation of people in advance of disastrous events). In most cases, such developments will remain under constant refinement in order to accommodate changes that will occur in different application areas. Hence, how to continuously apply new theoryes, new methods and new technologies of information science and computer science to service hydroinformatics is an opportunity and challenge for the development of water conservancy fields.

## 3. Model and Methodology

In order to better understand and represent the application of social media to hydroinformatics, a 4W (What, Why, When, hoW) model and a methodological structure is proposed in this section.

### 3.1 Hydroinformatics Model Based Social Media

The main research topics of hydroinformatics can be summarized as water quantity and water quality issues. On one hand, water quantity issue mainly focuses on the current total amount of water resources: too much water can lead to flooding while too little water will cause drought and drinking water scarcity. Therefore, hydroinformatics should apply the latest research methods and tools to provide comprehensive data and decision-making service for water resources monitoring and regulating, flood/drought predicting and disaster response. On the other hand, water quality is mainly concerned with whether water can be fit for human society, which corresponding to water environment monitoring and management in hydroinformatics.

As a latest technology, social media can provide basic data supports and assist decision-making assistances for hydroinformatics. In order to better represent the application of social media to hydroinformatics, this paper proposes a 4W (What, Why, When, hoW) model (Fig. 1), which expressed as follows:

1) What. The most important element of the proposed model is to answer what social media can provide for hydro-

informatics. As a massive and widely disseminated data source, social media can provide real-time, multi-attribute information and deep, assistant decision-making services for hydroinformatics in water quantity and water quality issues.

2) When. Social media has real-time temporal and spatial features. Therefore, social media can provide timely information for water-related business. And hydroinformatics can establish a pre-event, in-event, post-event decision-making and response services based on data and service provided of social media.

    3) Why. Traditional monitoring data, such as sensor data and environmental monitoring data, are not necessarily able to provide real-time and stable business monitoring data due to the restriction of monitoring site distribution and monitoring frequency. As a new type of crowd-sourcing data, social media can provide real-time, multi-attribute auxiliary data for different water business applications.

    4) HoW. According to the time dimension, social media can provide data and assist decision support for hydroinformatics at different stages of the water-related business, which including hydrological element monitoring, disaster prediction and early warning (pre-event); information sharing, communication and emergency response (in-event);event cause analysis, disaster relief and reconstruct(post-event).

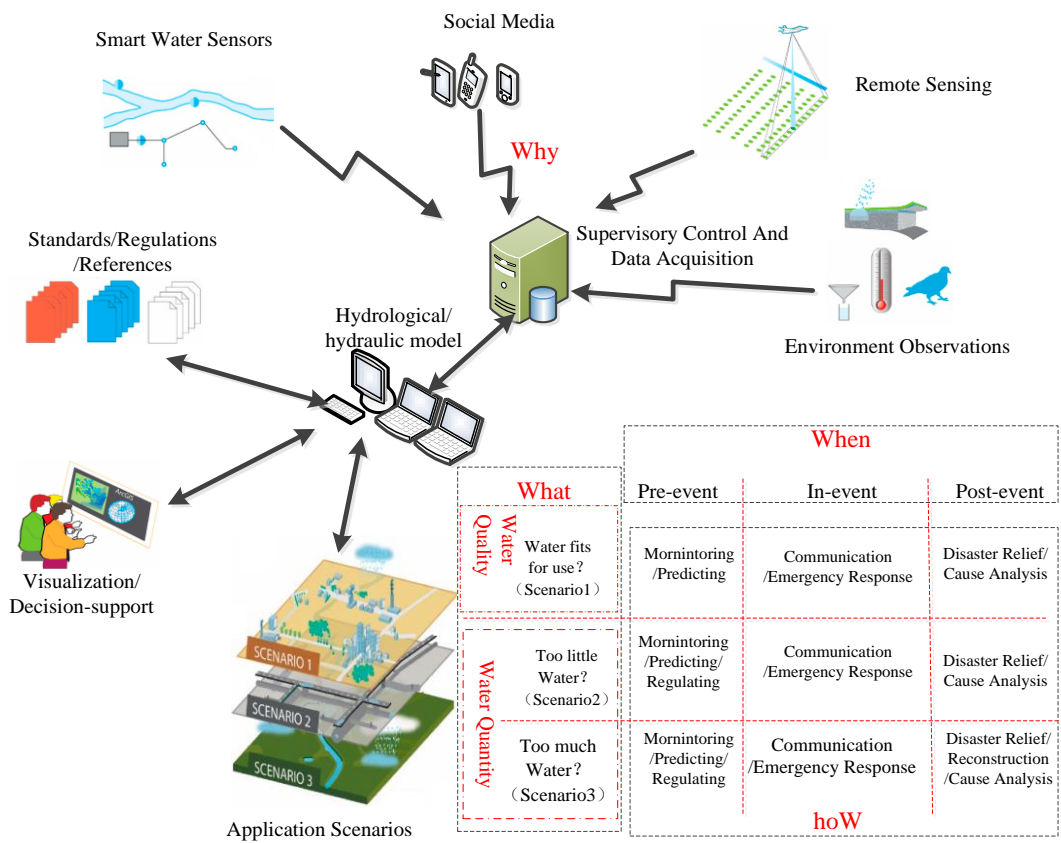

Figure 1: 4W model of apply social media to hydroinformatics.

## 3.2 Methodology

Fig. 2 displays the methodological structure adopted to apply data from social media into hydroinformatics. The methodology is divided into three stages: (1) hydrological data calibration and modelling (2) social media data transformation and modelling (3) comparison with real data. In each stage, a series of activities is carried out. Each of these processes is in turn explained in the next sections.

# APPLICATIONS OF SOCIAL MEDIA IN HYDROINFORMATICS: A SURVEY

(1) Hydrological data calibration and modelling

The first methodological procedure carried out was the calibration of the hydrological model that was used to obtain a transformation of authoritative and social media hydrological elements into other elements. It is a classic procedure in hydrology to use hydrometeorological variables such as rainfall and streamflow to calibrate the model [55]. In view of the fact that the methodology is designed to be used in ungauged and poorly gauged catchments or when there are sensors subject to failures, simple modelling seems to be more appropriate [56].

Transformation of authoritative hydrological data depends on the calibration performed. For example, the rainfall from authoritative gauges can be used to model the streamflow in the same period of social media harvesting. The simulated streamflow will be later compared with the one obtained from the social media modelling and the real values from authoritative sources. Low performance in calibration and validation is probably due to problems in the rain gauges, as already mentioned [57].

(2) Social media data transformation function and modelling

It can collect social media data by means of an API to fitting the transformation function. Following this, the messages are filtered by some filter such as geotag and keywords. As a result, the frequency of keywords is obtained and the variables are created. Then, an n-fold cross validation procedure for the fitting of the function is applied to regress the authoritative hydrological elements against social media data. In this procedure, a fixed time interval is removed from the sample and used later to validate the transformation function of the same time, and avoid any bias in the resulting function. These stages are repeated to obtain a transformation function for each month.

In transforming the social media data into hydrological elements, data were collected inside the catchment to obtain this element for this place. It should collect the same variables with the same temporal resolution examined. Once the tweets had been collected, the frequencies of the tweets were replaced inside the function created in the past section. However, since hydrological processes, like rainfall-runoff, are only possible in systems such as catchments, where the boundaries do not necessarily match the administrative boundaries of the city, a "regionalization" of the tweets within a catchment-area is carried out by dividing the frequencies of the related tweets every 10 min within the drainage area of the catchment. Thus, this process differs from the parameter fitting process. Finally, the estimated hydrological values were used as input of the hydrological model to generate the other hydrological elements.

(3) Comparison of the joint use of traditional hydrological modelling and modelling from social media.

This step involves comparing real hydrological data, with estimated values calculated from social media messages and authoritative hydrological modelling. This comparison is made by determining if the real hydrological data are found within the confidence interval of the models, or have been overestimated /underestimated instead. This assessment makes it possible to establish the accuracy of these cases when the modelling is only carried out by means of social networks data, and employing the transformation function to estimate hydrological data for the "ungauged" catchments, i.e. when we do not have to rely on authoritative sensors. Additionally, we analyzed the case when the results from both models are employed, by selecting the maximum and minimum values of the confidence interval of each model and evaluating their accuracy to predict real streamflow values. This scenario is equivalent to the case of "poorly gauged" catchments, where data from both sources is available but the authoritative data are inaccurate and/ or imprecise.

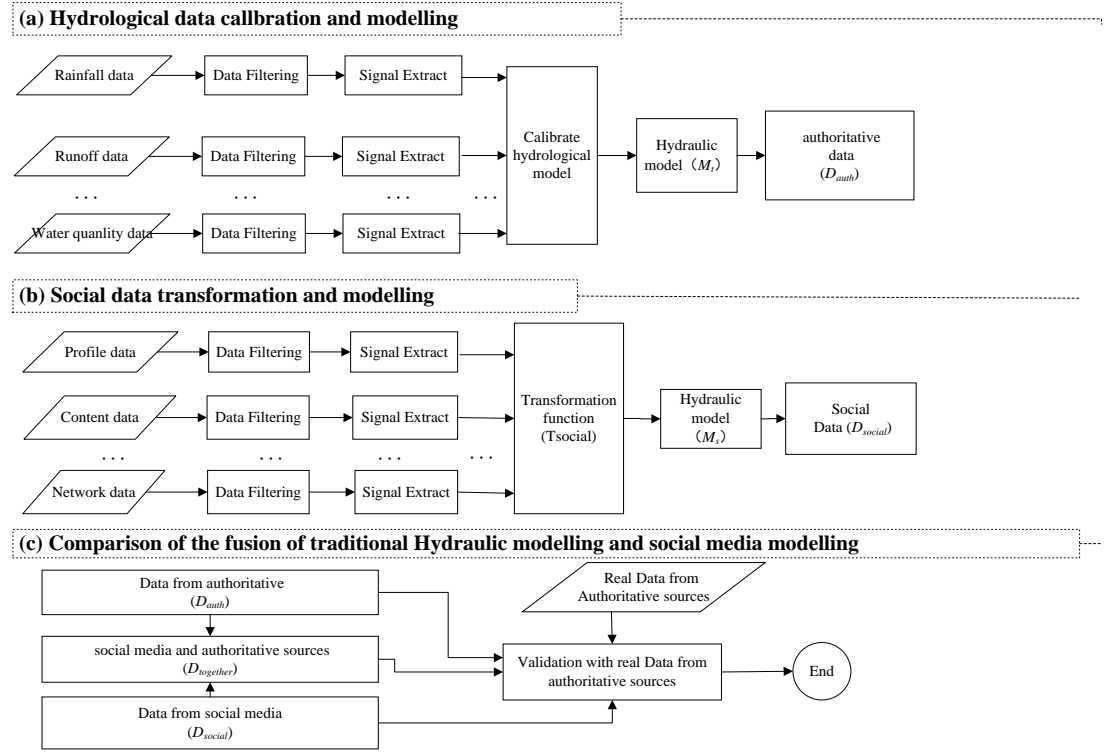

Figure 2: Methodological structure to fusion social media in hydroinformatics.

## 4. Applications of Social Media in Hydroinformatics

Social media has played an active role both in water quantity and water quality issues. In this section an overview of academic research of applying social media to hydroinformatics such as water environment, water resources, flood, drought and water scarcity management is provided.

### 4.1 Social media applications in Water quantity Issues

Water is a critical natural resource that has significant impacts on human living and society. Growing population and energy consumption exacerbate the quantity of water and our ability to manage this resource. Water quantity is a measure of the amount (e.g., volume or discharge) of water supplied by a contributing watershed. The relative amount of water available in support of ecosystem services depends on the quantity of water delivered to a landscape and how it is partitioned within the landscape for urban, agricultural, industrial, conservation, and other uses.

Water quantity is one of the most pressing management and environmental issues. A continually increasing population along with ever-intensive, irrigated agriculture continues to increase demands for water [58], and subsequently affect natural water conveyance hydro-periods. Applying social media in water quantity management include three main aspects: daily water

# APPLICATIONS OF SOCIAL MEDIA IN HYDROINFORMATICS: A SURVEY

resources management (normal water resources), flood forecasting and management (too much water resources), and water scarcity management (too few water resources).

**4.1.1 Water Resources Monitoring and Management**

Water resource problems are extremely complex due to their scope, scale, and interconnection between multiple systems crossing diverse disciplinary and social boundaries. Several problems arise with water supply, usage, conservation, and treatment restraint [59]. However, in a realistic operational environment, these operations are run and managed under different institutions and business entities in an isolated and independent manner. The amount of water that vendors pump depends heavily on the market need, which results in an unbalanced water supply and demand. Furthermore, residential areas need guidance if they are to adopt more economical habits when consuming water. Due to the chaotic situation, it is highly valuable to build a smart and connected water platform for daily water resource management.

As an important part of "Europe2020"[60] funded project, SmartH2O[61] aims at creating a virtuous feedback cycle between water users and the utilities, providing users' information on their consumption in quasi real time, and thus enabling water utilities to plan and implement strategies to reduce/ reallocate water consumption. It developed an Information Communications Technology platform to design, develop and implement better water management policies using innovative metering, social media and pricing mechanisms. Planned case studies in the UK and Switzerland showed that high quality data obtained from smart meters and communicable through social media could be used to design and implement innovative and effective water pricing policies.

The Murray-Darling Basin Authority in Australia [62] developed a project to use social media for connecting communities to the organization and conversational control. They used social media as a tool to influence the flow of conversation, to enhance communication between stakeholders and to meet the needs of the community. The data has indicated that human relationships and communication between community members is enabled through the use of social media.

Facing the increased strains to water and wastewater infrastructure, our cities need to change and develop a smart way to accommodate all that growth and make our cities "smart cities". The urban water sector of Europe [63] established a good understanding of needs and concerns of stakeholders at the local regional and river basin level, and efficient management and governance of water-related problems. They built Transport and Information & Communication Technologies tools to influence human behavior, alter social patterns, to inform a wide audience about the significance of water in the cities of the future, and thus helping citizens and stakeholders to develop sustainable habits regarding urban water use through social media and gaming.

Not only does the water administration increasingly rely on social media assistance in developing policies related to water management, the public often turns to social media for assistance when dealing with daily water-related issues. A case in point is the domestic water charges in Ireland. Martin et.al [64] explores water governance and stakeholder engagement during the introduction of domestic water charges in Ireland through a Twitter dataset. The findings provide some insights into the role of social media in water governance and stakeholder engagement issues in an Irish and wider context. Nguyen et .al[59] developed a WaterScope prototype platform to collect, integrate, store and manage a variety of dynamic and heterogeneous water-related datasets such as individual water level data, weather, social media data, and water knowledgebase data resources. Furthermore, the tool enables forecasting underground water levels, identifying water concerns, sharing knowledge and expertise among stakeholders, and thus bringing new insights to our understanding and insights of the water supplies and resource management.

In short, the use of social media in water resource management area fits in and meets the water challenges and objectives as it can support water utilities in determining optimal water pricing and consumers in chancing their water consumption habits, thus dually contributing to the target of a more efficient use of water. This is achieved by integrating smart metering, social computation, dynamic water pricing, and advanced consumer behavioral models. Furthermore, it highlights the importance of innovation in the water sector by coupling smart meter technologies with innovative end-user services which can reach better water management through the means of rewards, automation and information.

**4.1.2 Flood Monitoring and Management**

Flood is one of the most widespread types of natural disasters with a wide range of influences, long duration, and large losses in the world [65-66]. With the influence of urbanization, deforestation, subsidence and climate change, the temporal and spatial distribution of precipitation will be more uneven, which will lead to frequent flooding event.

However, due to complex nature environments, low flood control standards and lack of data supporting, the damage of small and medium-sized rivers floods and urban floods accounted for 70-80% of the total damage, and will show increasingly frequent trends [67]. For this reason, flood forecasting, monitoring and management have attracted a great deal of attention as a means of improving early warning systems [68-70]. Social media can provide forecasting and monitoring,response and communication, flood damage assessment and disaster relieving services for flood management at pre-even, in-even and post-even stages.

Flood forecasting and monitoring have attracted a great deal of attention as a means of improving flood early warning systems [71]. It integrates hydrological, meteorological and underlying surface information to forecast floods and processes such as future runoff, water level according to hydrological laws. A reliable, robust and efficient flood early warning system can provide scientific decision support for flood management and relief [70]. Flood forecasting and monitoring are being increasingly characterized as a problem of "big data", since there are different data sources that can be used to support decision making, such as satellites, radar systems, rainfall gauges and hydrological networks [72]. However, in the pre-event of a flood, the apparent overabundance of data is often accompanied by a simultaneous "information dearth": a lack of information may arise because sensors are not available for certain regions or the number of available sensors is not enough to cover the territory with a suitable resolution.

The advance of mobile telecommunications and the widespread use of smartphones and tablets allow people to act as human sensors, and generate Volunteered Geographic Information (VGI) [73]. Moreover, the extensive spatial coverage of the measurements monitoring by social

# APPLICATIONS OF SOCIAL MEDIA IN HYDROINFORMATICS: A SURVEY

media makes it possible to obtain useful information at different points of river catchment areas and cities where the local inhabitants are able to supplement the static sensors of the hydrometerorological networks. Thus, social media have been increasingly recognized and used as an important resource to support flood early warning systems [74]. Camilo et.al[57] found that there were close spatiotemporal links between social media activity and flood-related events [75] as well as social media activity and rainfall [76]. They use relevant information extracted from social media and hydro-meteorological sensors in streamflow modelling to predict streamflow and flood conditions, to assist in issuing early flood warnings and to improve rainfall run-off from observational, authoritative networks and even observed urban streamflow; evidence showed that better results can be achieved by merging authoritative data with information from social media. Wang et.al[77] employed social media and crowdsourcing data to complement the datasets developed based on traditional remote sensing and witness reports, which will help to address the issue that unable to analysis flood risk, control flood disaster and validate hyper-resolution flood model caused by lock of datasets for urban flooding.

In times of mass emergencies, collective behaviors that include intensified information search and information contagion apparent [78]. During the flood, people want to know where exactly their families and friends are as not being able to reach them or knowing they might not be able to contact you can be very frightening moments. Thus, information become critical as the availability of immediate information can save lives. People share information about real-time flood situation, approaching threats, where to evacuate, where to go for help, etc. Not only do they want to know about the destruction that has occurred, but the government and nonprofit organizations also eager to help those affected by searching for victims, providing relief supplies and raise funds from donations. Thus, there is a need to keep abreast of the latest developments, however, this is difficult since information produced under crisis situations is usually scattered and of varying quality.

Social media is enabled by communication technologies such as the web and smartphones, and turn communication into an interactive dialogue to provide the necessary breadth and immediacy of information required in times of emergencies. It can offer a unique, fast and effective way to disseminate information about individuals, associations and government of flood events [79-80]. Cheong et.al [79] uses Social Network Analysis (SNA) to study interaction between Twitter users during the Australian 2010-2011 floods. They developed an understanding of the online community that was active during that period to find the online social behaviors, the influential members and the important resources being referred to. The result indicates that social media plays positive roles in flood management and can be reasonable to push for greater adoption of social media from local and federal authorities Australia-wide during periods of mass emergencies.

Besides information sharing and communicating, many previous works in this area have concentrated on using social media data either for rapid flood inundation mapping [81-82] or flood risk management [83-85]. Rapid flood mapping is crucial for flood disaster response to gain better situation awareness during the event, and thus to quickly identify areas needing immediate attention. Li et.al [81] introduces a novel approach to mapping the flood in near real-time by leveraging Twitter data in geospatial processes. They developed a kernel-based flood mapping model to map the flooding possibility for the study area based on the water height points derived from tweets and stream gauges, and then used the identified patterns of Twitter activity to assign the weights of flood model parameters. Smith et.al [83] presents a real-time modelling framework to identify areas likely to have flooded using data obtained only through social media. The framework demonstrates that social media provides an excellent source of data, and that its utility may be further enhanced when coupled with efficient GPU accelerated high-resolution hydrodynamic modelling.

When flood occurred, it often causes more and more serious damage to our human's daily affairs such as food, water, sanitation and shelter. Thus, how to quickly carry out flood damage assessment and post-disaster treatment/ reconstruction is also an important part of flood management. In recent years, governments and relief organizations are increasingly utilizing the power of social media to access information tools such as flood damage assessment [86-88] and social media-derived relief efforts [89-90]. Brouwer et.al [86] investigates the December 2015 flood in the city of York (UK), and then presents and evaluates a method to create deterministic and probabilistic flood damage extents from Twitter messages that mention locations of flooding. This study illustrates that social media content has real potential in generating flood extent estimates and therefore can be used to gain insight into the current situation of flooding. Chong et.al [89] provides a detailed examination of grassroots uses of social media aimed at soliciting disaster -related assistance during the Kuantan (Malaysia) Flood. This research gives meaningful exploration and guidance to the role of social media in flood relief and post-disaster reconstruction.

Currently, social media has played an active role in flood monitoring, information sharing and disaster response during flood, disaster assessment and post-disaster reconstruction. However, some issues such as how to improve the accuracy of flood forecasting, how to control rumors in disaster, are still lack relevant research and need further scientific work.

**4.1.3 Drought and Water Scarcity Management**

Drought is distinct, natural calamity because it is a slow-onset, creeping phenomenon that can cause long-lasting and wide-ranging impacts [91]. Drought occurs when water resources cannot meet demands over most parts of the world. In general sense, it does not have a sudden beginning or a clear end and results from a deficiency of preci-
pitation over an extended period of time (usually a season or more) and thus lead to water shortage for some activety, group, or environmental sector. Drought is not necessarily very visible and not necessarily causes structural damage to infrastructure, but it is still ranked as a very severe hazard because it can spread over a large geographical area and cause serious water Scarcity[92-93].

Compared to other hazards, there is limited research has focused on applying social media to analyze the drought risk communication and management. Sonnett et al.[94] analyzed drought-related information published on main newspapers and identified how discursive context can shape the framing of drought in temporal and spatial scales. Mass media can serve as a valuable mechanism to deliver information regarding drought risks and impacts to citizens, thus improving the levels of awareness on the sustainable use of water resources [95]. Tang et al. [96]

highlighted that spreading information via social networks, rather than via the traditional mass media, is important to enhance awareness and perception on water conservations. Tang et.al [97] discussed use of social media and VGI in the management of the 2014's California drought [98]. They proposed SWOT (Strengths, Weaknesses, Opportunities, and Threats) model to evaluate the social media sites of governmental agencies that were directly involved in California's Drought Task Force during drought. The results show that the popular social media platforms (Facebook, YouTube, and Twitter) have been used as an efficient communication channels between professional stakeholders and the general public on delivering first-hand information, preparing and training people in combating processes in the whole practice, and thus play a significant role in drought mitigation and management.

Wagler et.al [99] used social media monitoring and analysis to explore online Twitter conversations related to 2012-2013 historic drought in Nebraska during a one-year period. The study found that social media served as a news outlet for information and updates about drought conditions, its conversations such as agricultural issues, environmental impact, extreme weather, effects on the public, and proposals of solutions to address drought increased in quantity as drought conditions worsened. It also suggested that educational institutions and organizations should serve as leaders on social media and in social networks to disseminate timely and relevant information related to important public issues, while also monitoring and participating in surrounding discussions.

Vaishali et.al [100] built a Naïve Bayes multi-label classifier to provide topic recommendation, and human analysis results for big data mining dataset from social media during drought 2016 in India. The author claimed that social media data mining can provide great help for relevance decision makers to gain further understanding of drought impact and action for rehabilitation of drought affected area by government agencies, NGO and private organization.

When drought occurred, it is likely to cause a serious water scarcity. Thus, how to use social media to response and address the crisis caused by water scarcity attached a lot of research attention. Pettersson et .al [101] used social media as a tools to understand why people turn to social media in a crisis and analyses whether different types of users resort to social media during a crisis for different reasons in water Scarcity of Cape Town, South Africa. They used the Different Users and Usage Framework [102] to obtain information, then applied the Theory of Planned Behavior [103] to assist on explaining three main findings (1) People turn to social media during a crisis for different reasons (2) According to the analytical results, different users tend to dominate different usage areas and (3) During the Cape Town water Scarcity, it was common practice for businesses and corporations to raise awareness and combine it with promoting their business.

As a new information acquisition and update channels, social media can play an important roles in drought management includes one and two-way information sharing, situational awareness, and decision- making. Though some people have made useful attempts and explorations to apply social media in drought management, many major studies such as two-way communication, reconnection between public social media domain and personal social networks and rumor control still stayed in relatively surficial levels with limited data sources and empirical study, which will be future research direction.

**4.2 Social media applications in Water Quality Issues**

Water quality is a measure of the chemicals, pathogens, nutrients, salts, and sediments in surface and groundwater. Not only is water quality important for drinking water supplies, but quality is an important attribute of all other water provision services, such as production of fish and other fresh-water organisms that are consumed by humans. Surface freshwater is a finite resource that is necessary to the survival of mankind and the ecosystem. Adequate quantity and quality of water are also essential for sustainable development [104]. However, many surface water systems have been contaminated by treated or untreated wastewater that has been discharged by domestic, industrial, and agricultural water users. Water quality has also become an important component of the global water crisis.

**4.2.1 Water Quality Monitoring and Protection**

Surface water quality monitoring (SWQM) provides essential information for water environmental protection. A water quality monitoring network requires monitoring sites, frequency, variables, and instruments as well as trained/educated field personnel. However, establishing a SWQM network in a broad area entails huge costs [105].Compared to traditional environment monitoring methods utilizing expensive and complex instruments, social media analysis is an efficient and feasible alternative to achieve this goal with the phenomenon that a growing number of people post their comments and feelings about their living environment on social media, such as blogs and personal websites.

Zheng et.al [106] provides a framework for collecting water quality data from citizens and offers a primary foundation for big data analysis in future. They built Tsinghua Environment Monitoring Platform(TEMP, http:// www.thuhjjc.com) based on WeChat, through which TEMP users can describe and take photos of river and lake waters, report the surface water pollution activities that affect their living and health following the TEMP instructions. Reports based on water quality and water pollution information collected from TEMP indicate that the citizen-based water quality data are relatively credible if the volunteers are trained in water quality monitoring.

Wang et.al [107] self-defined a term called the Environmental Quality Index (EQI) which including water quality, food quality and air quality to measure and represent people's overall attitude and sentiment towards an area's environmental quality at a specific time, and then constructed a new environment evaluation model to monitor environmental quality by calculating and analyzing the EQI collected from social media. The experiments results on 27 provinces in China in 2015 by utilizing this environment evaluation model show that the environment evaluation model constructed based on social media is feasible. Furthermore, this research provides a foundation for monitoring environmental quality by analyzing social media information.

Considered that VGI and social networking in WebGIS has the potential to increase public participation in soil and water conservation, promote environmental awareness and change, and provide timely data that may be otherwise unavailable to policymakers in soil and water conservation management. Werts et.al [108] developed an integrated framework for combining

# APPLICATIONS OF SOCIAL MEDIA IN HYDROINFORMATICS: A SURVEY

current WebGIS technologies, data sources, and social media for soil and water conserveation. The experiments on sediment pollution of abandoned developments in upstate South Carolina indicated that the use of social media does not replace the need for expert opinion and analysis in soil and water conservation but may allow a small number of experts to efficiently complete initial evaluations of a large number of locations.

**4.2.2 Water Contamination Crisis Response and Treatment**

The natural water cycling process provides critical products and services like fresh water, purification and recreation, based on which human civilization has been sustained [109-111]. However, it has now become clear that our humans' presence, behaviors and their consequences have begun to affect the earth's hydrology and thus led to the serious water scarcity, the widespread contamination of water, the soil loss in food production areas and rapidly expanding globally.

Water contamination is the most important environmental problem at current society. Water contamination crisis worldwide are becoming increasingly fierce and the challenges for sustainable water resource management and society development [112]. The global water contamination crisis has not only brought about devastating human suffering such as diarrhea, cholera, and various skin diseases, but also brought environmental damage, economic losses, and social impact. In its 2015 annual risk report, the World Economic Forum lists water contamination crises as the largest global risk in terms of potential impact [113]. Therefore, how to scientifically and reasonably handle water contamination crisis is a key issue in the application of water quality management.

In recently years, the importance of social media as a vital tool has increased in the crisis management and crisis response. There have been many studies done on social media from the audience perspective on why/how they use social media in a water contamination crisis, how they judge credibility of sources and information, how organizations are embracing Web 2.0/ new media and incorporating it into their crisis planning and response [114-116]. Generally speaking, crisis management comprises three stages: crisis prevention, crisis response, and crisis recovery. Albala-Bertrand et.al [114] suggested that we should pay more attention to the crisis response stage in order to reduce and absorb the effects of disaster as they occur because crisis prevention can be futile given that crises are usually unexpected[115]. Moreover, effective crisis response can lessen crises' indirect impact, and may help to "reverse the direct effects" in crisis recovery [115]. Getchell et.al [116] provided a network analysis of official Twitter accounts activated during the Charleston, West Virginia, water contamination crisis in 2014. They built a social media network using 41 official Twitter accounts from people and organizations directly related to the water contamination, then used social media data mining tools to analysis and examine the structure of the network formed by these Twitter accounts. Though this analysis only scratches the surface of the potential for applying social network analysis methods to risk and crisis communication, but it reveals that new media are changing the face of risk and crisis communication both in the ways that the public seeks information and messages for self-protection from official sources and the way organizations involved in the incident communicate and share information with each other and other stakeholder groups.

Lanham et.al [117] conducts in-depth research as well as surveys on various social media posts to study real life activism versus passive in regards to the Flint, Michigan water crisis [118]. The survey analyzes how social media posts raise public concerns about environmental crises and how government officials use social media to monitor civic activities and discuss crises. The results reveals that social media had a large impact on people's knowledge of environmental issues and can be used to determine if social media is an effective outlet to promote future environmental crises and the activism vs. slacktivism issue.

Apart from this, there are many other ways social media can be used in a crisis situation to 'strengthen situation awareness" and improve emergency response [105-108]. From individual's perspective, through following official natural disaster management agencies on social media, ordinary social media users can be alerted to authoritative situational announcements; From an organizational perspective, disaster response organizations can leverage social media as a platform to communicate with the public in disaster situations and potentially solicit on-the-ground information using the public as information sources. Furthermore, the ability to monitor citizens' opinions can keep the government and organizations updated, as well as the fact that the public are often some of the first responders during a crisis. Besides, when authority actors post information about a crisis on social media, it can contribute to calm the public. When people are better involved and informed about a crisis they are more likely to take on an optimistic approach.

## 5 Research Trend and Open Problems

With the large number and rapid growth of social media systems and applications, social big data has become an important topic in a broad array of water conservancy research areas. The purpose of this study is to provide a holistic view and insight for potentially helping to help find the most relevant solutions that are currently available for applying social media mining techniques to Hydroinformatics.

As such, we have investigated the state-of-the-art technologies and applications for servicing social media big data in hydroinformatics. These technologies and applications were discussed in the following aspects: (i) what are the main representative research issues in social media mining? (ii) How does one mine social big data to discover meaningful patterns? and (iii) How can these patterns be exploited as smart, useful user services through the currently deployed examples in water-related applications and hydroinformatics?

More practically, this survey shows and describes a number of existing researches (e.g., flood prediction and management，water crisis response and treatment) that have been developed and that are currently being used in water conservancy based on social media. Although it is extremely difficult to predict which of the different issues studied in this work will be the next "trending topic" in water-related social big data research, from among all of the problems and topics that are currently under study in different areas, we selected some "open topics" related to data collection, data quality management, fake news detection, security and privacy issues, analysis algorithms and platforms, providing some insights and possible future research.

### 5.1 Data Collection

One of the most important criteria for scientific research is the falsifiability of hypotheses and

# APPLICATIONS OF SOCIAL MEDIA IN HYDROINFORMATICS: A SURVEY

scientific theories[119]. Scientists can use the same methods or data from previous publications to validate or replicate published research results. Although modern telecommunication facilities provide easy setup of data transmission for the efficient and centralized data collection and handling for water network installations all around the world, there is still facing serious problems in data collection for applying social media in hydroinformatics. On the one hand, as one of the branch of big data, water resource data have their own 4V feature including volume, variety, veracity, and velocity. Thus, to fuse, process and analysis those structured and unstructured big data that collected by multi-source water network sensors on daily basis still remains challenging for several reasons: the difficulty to manage distinct protocols for data collection; the need to integrate data sets of different provenance, coverage, granularity and complexity; the time constraints of a water-related application context [2]. On the other hand, though social media big data is accessible through APIs, there are still very few affordable data provided to academia and researchers by social data sources due to the commercial reason [120]. For example, news services such as Thomson Reuters and Bloomberg typically charge a premium for access to their data. Though Twitter has recently announced the Twitter Data Grants program, where researchers can apply to get access to twitter's public tweets and historical data in order to get insights from its massive set of data, researchers can only retrieve 1% of randomly sampled Twitter data (tweets) via public APIs for their research due to the reason that researchers cannot re-distribute the original "raw tweets" collected in their databases to others except for their internal research groups [119]. Therefore, researchers can hardly re-test, re-run or further analysis recent published social media big data research. All in all, these data sharing and scraping issues may hinder the development of big data and social media research for hydroinformatics in the future, and it is also one of the research hotspot of social media big data mining and application [121].

**5.2 Data Quality Management**

By rapidly acquiring social media big data from various sources, researchers and decision-makers have gradually realized that this massive amount of information has benefits for understanding customer needs, improving service quality, improving prediction accuracy and reducing decision risk. High-quality data are the precondition for analyzing and using social media big data and for guaranteeing the value of the data. However, the uncertain and unstructured nature of this kind of big data presents a new kind of challenge: how to manage the value of data and evaluate the quality of data.

Data quality is a process of assessment of data. It can be defined as data that are fit for use by data consumers [122]. Hydrological data is an important source for many applications in water-related engineering and widely used for designing storage reservoirs, flood protection measures or for prediction purposes. Therefore, ensuring the quality of hydrological data, especially data quality issues regarding social media data have been highlighted as one of the grand challenges for the development of hydroinformatics in the social media era. [123-125].

There will be many reasons for the low quality of water-related social media data: information incomplete, information inconsistency, information irrelevancy, information out-of-date and information unbelievable [126]. Hence, to address above-mentioned data quality problem and extract high-quality and real data from the massive, variable, and complicated water-related social media big data sets becomes an urgent issue due to following challenges [127-128]:

1) The diversity of data sources brings abundant data types and complex data structures and increases the difficulty of data integration.

2) The tremendous of data volume makes it difficult to judge data quality within a reasonable amount of time.

3) The rapid change frequency and short "timeliness" necessitates higher requirements for processing technology.

4) The fact that no unified and approved data quality standards have been formed in the world makes it more difficult to form a comprehensive and universal data quality assessment and management system.

**5.3 Fake News Detection**

Social media for news consumption is a double-edged sword. On the one hand, its low cost, easy access, and rapid dissemination of information lead people to seek out and consume news from social media. On the other hand, it enables the wide spread of "fake news", i.e., low quality news with intentionally false information [129]. The extensive spread of fake news has the potential for extremely negative impacts on individuals and society. It has affected stock markets [130], slowed responses during disasters [131], and terrorist attacks [132]. Recent surveys have alarmingly shown that people increasingly get their news from social media than from traditional news sources [133-134], making it of paramount importance to curtail fake information on such platforms [135]. According to the World Economic Forum [135], the wide impact of fake information makes it one of the modern dangers to society due to the motives of influencing opinions and earning money [136].Therefore, fake news detection on social media has recently become an emerging research that is attracting tremendous attention.

Social media platforms have allowed individuals and organizations to share information with their peers and specific audiences. Information typically is shared with good intent. However, Due to lack of access to actual, authoritative and real-time information, humans are susceptible to accept and spread the latest, unreleased information which may include rumors, fake information and misinformation (e.g., deception, propaganda and malicious spamming) on social media, especially proliferate before, during and after disasters and emergencies[137]. The latest research shows that social media has played a negative role in fueling false news dissemination [138]. For example, during 2018 Kerala floods disaster in India, floods of fake news on social media created unnecessary panic [139] and fueled confusion and fear [140]. Fake news also has been found at water scarcity crisis of Cape Town, South Africa [101,141] and drinking water crisis in Lake Taihu, China [142-143].

As it well known, characteristics of fake information may include uncertainty in the "facts," emotional exploitation of a situation, trending topic discussions for hijacking conversations and financial scams, among others. Though this information cannot be completely eliminated, different research efforts such as early fake news detection, first responder agencies can use various tactics and strategies to give early alerts of fake news during the dissemination process and thus to offset bad information. However, detecting fake news on social media is quite challenging because it is written to intentionally mislead users, and attempt to distort truth with different styles while mimicking real news, which makes existing detection algorithms from traditional news media ineffective or not applicable.

Another biggest challenges public safety agencies and organizations face is how to reduce or eliminate the spread of fake information, especially as public demands for a response from these authorities' increases. Social media can distribute news including misinformation, false information and rumors faster and to a wider audience than traditional news sources. Therefore, how to increase public trust in government, media and nongovernmental organizations (NGOs) so as to improve the ability of first responders to mitigate and minimize false information, fake information and rumor spread, is a challenging task for water-related social media mining.

## 5.4 Security and Privacy Issues

The issue of water security-defined as an acceptable level of water-related risks to humans and ecosystems, coupled with the availability of water of sufficient quantity and quality to support livelihoods, national security, human health, and ecosystem services [144-145]-is thus receiving considerable attention. To this end, it is necessary to establish sound information security management systems accordance with relevant information security policies and regulations, and conduct effective data mining and data application research without exposing confidential information.

On the one hand, water-related data is a key piece of information for water security and national security. Therefore, it should establish a unified water-related social media big data security hierarchy and system to strengthen the security protection and supervision capabilities of important water-related data and infrastructure networks. Moreover, it is necessary to set up a secure and reliable data transmission, migration and delivery mechanism to prevent data from being compromised and stolen as well as a rights management and audit mechanism to ensure that the data is correctly used and managed.

On the other hand，flooding of user-generated social media data (including links, posts, and profile information) increase the risk of exposing data security and personal privacy as it is rich in content and relationship and contains individuals' sensitive information, which will lead to an increasing risk of privacy breaches [146-148]. Therefore, it should adopt some strategies by anonymizing or removing "Personally Identifiable Information" like user names, ID, age and location information and keep the social graph structure to protect social media data publishers' privacy [149]. However, this solution has been shown to be far from sufficient to protect people's privacy [147]. Consequently, various protection techniques have been proposed for anonymizing each aspect of the heterogeneous social media data such as graph data structure [150], and users' textual information [151].Moreover, how to communicate secure data, match graph and identify social when social data are merged from available sources ,how to build useful benchmark datasets to evaluate and test privacy-preserving services, are still open issues and potential research areas related to privacy[152].

### 5.5 Social Media Analysis Algorithms and Platforms

With the advancement of hydro-informatics and hydro-modernization, water-related data has greatly expanded in terms of time and space scales and element types. Using social media big data analysis technology and mining tools to explore the hidden pattern and interrelationships between water-related social media big data, and then to express this knowledge through data visualization, so as to provide decision support for flood prediction and management , water resources plan and management, water crisis response and treatment. General speaking, design adequate approaches to process, analyzing massive amounts of online data are the crucial step for water-related social media big data analysis. However, the research and application of water-related social media big data analysis and mining algorithms are still in the beginning .Though there has some prior research on social media analysis methods related to hydrological data acquisition, flood control and management, there are still a long way to adopt sophisticated differential equations, heuristics, statistical discriminators (e.g., hidden Markov models), and artificial intelligence machine learning techniques (e.g., neural networks, genetic algorithms and support vector machines) to identify and extract subjective information from water-related social media big data.

Moreover, analysis and decision-making services related to water conservancy often require statistical data, but current decision-making analysis systems often need to query multiple information systems and external systems based on various heterogeneous data sources, and then perform a large amount of data analysis [153]. This mode of work has a large workload, low data utilization, and is prone to errors. In this respect, it is urgent to build a comprehensive social media big data decision-making analysis platform which including data storage and sharing, data security, data mining and data visualization to effectively reduce decision-making risks and improve decision-making efficiency. Hence, it should apply the latest computer and information technology such as service-oriented architecture [154] and cloud computing [155], service encapsulation and composition technology to meet the changing water conservancy requirements of business application and collaborative management. It also should integrate GIS, workflow engine, visualization tools, etc. to provide dynamic and intuitive support services for water-related departments, companies and public services.

### 6 Conclusion

The social media era is an upcoming trend that no one can escape from. The concept of social media originated from the popularization of Web2.0 as digitalizing of the information among the world becomes much easier and cheaper for future data mining purpose. The idea of social media and big data is very adaptable, and can be valuable for academic purpose as well. Scientists are expected to embrace the big data era rationally without being blurred by the overwhelming trend.

Water-related social media big data is the trend of water science development, and also an important application area of social media and big data research. On the one hand, with the continuous improvement of hydroinformatics, crowdsourcing, web-based and mobile-based ICTs, the water authority has accumulated huge volume distributed multi-source and heterogeneous water-related business/social media big data .Therefore, using social media tools and technologies to process , analyze and mine useful and meaningful information from those data can benefit to 1)

# APPLICATIONS OF SOCIAL MEDIA IN HYDROINFORMATICS: A SURVEY

analyze the climate change, flood and drought management and the global water cycle 2) promote harmonious coexistence between human society and water environment and 3) optimize the water resource planning and management systems. On the other hand, Social media big data is a highly integrated science and its theoretical system is still developing and progressing. So, water-related researchers should learn and use the latest big data as well as social media mining technologies to improve the management decision-making level of the water industry.

At percent, the research and application of water-related social media big data are still in the initial stage. It is necessary to formulate unified industry standards to promote the application of social media big data in water- related business.

The application and research of social media big data in hydroinformatics is a complex project. This paper has conducted an in-depth research and analysis on water-related social media application. The paper first explained the background of water-related social media big data, and then introduces the basic theory of social media and social media mining, the application prospect of social media and the research status of hydroinformatics. After that, the paper systematically summarized the application and research methods of water-related social media big data in hydroinformatics core business such as flood forecasting and management, water resources monitoring and management, water environment monitoring and inclusion and water crisis response. Based on above research, the paper puts forward suggestions for the development of water-related social media big data from t data scraping, data quality management, fake news detection, privacy issues and social media analysis algorithms and platforms for managers and researchers.

**Acknowledgment**

This work has been partially supported by the CSC Scholarship, The National Key Research and Development Program of China (*Nos.2018YFC0407900*) and The Fundamental Research Funds for the Central University(HHU, 2018B45614).

# APPLICATIONS OF SOCIAL MEDIA IN HYDROINFORMATICS: A SURVEY